\documentclass [aps,prl,twocolumn,showpacs] {revtex4}
\usepackage[final]{graphics}
\usepackage{amssymb}
\usepackage{amsfonts}
\usepackage{epsfig}

\begin{document}

\title{Critical behavior of the spin-$3/2$ Blume-Capel model on a random 
two-dimensional lattice}

\author{F.W.S. Lima$^{1}$ and J. A. Plascak$^{2}$}
\affiliation{$^{1}$Departamento de F\'{\i}sica,
Universidade Federal do Piau\'{\i} , 57072-970, Teresina, PI, Brazil}
\affiliation{$^{2}$Departamento de F\'{\i}sica,
Universidade Federal de Minas Gerais, C. P. 702, 30123-970, Belo Horizonte, 
MG, Brazil}

\begin{abstract}

We investigate the critical properties of the spin-$3/2$ Blume-Capel model
in two dimensions on a random lattice with quenched connectivity
disorder. The disordered system is simulated by applying the 
cluster hybrid Monte Carlo update algorithm and re-weighting techniques.
We calculate the critical temperature as well as the critical point 
exponents $\gamma/\nu$, $\beta/\nu$, 
$\alpha/\nu$,  and $\nu$. We find that, contrary of what happens to
the spin-$1/2$ case,  this random 
system does not belong to the same universality
class as the regular two-dimensional ferromagnetic model. 

\end{abstract}


\maketitle

\section{Introduction}
Experimental studies of the critical behavior of real materials 
are often confronted with the influence of impurities and inhomogeneities \cite{bel}.
For a proper interpretation of the measurements it is, therefore, important
to develop a firm theoretical understanding of the effect of such random 
perturbations. In many situations the typical time scale of the thermal fluctuations in
the idealized ``pure" systems is clearly separated from the time scale of the impurity
dynamics, such that to a very good approximation the impurities can be
treated as quenched.
The importance of the effect of quenched random disorder on the critical
behavior of a physical system can be classified by
the specific heat exponent of the pure system, $\alpha_{pure}$.
The criterion due to Harris \cite{harri} asserts that for
$\alpha_{pure}>0$ quenched random disorder is a relevant perturbation,
leading to a different critical behavior than in the pure case
(which is the case of the three-dimensional Ising model). In
particular, one expects \cite{fisher} in the disordered system that
$\nu\geq 2/D$, where $\nu$ is the correlation length exponent and $D$ is the dimension 
of the system.
Assuming hyper-scaling to be valid, this implies $\alpha=2-D\nu\leq 0$. On the other hand
for $\alpha_{pure}<0$ disorder is irrelevant (as is the case of the three-dimensional
Heisenberg model) and, in the marginal case
$\alpha_{pure}=0$, no prediction can be made.
For the case of (non-critical) first-order phase transitions it is known that the influence of
quenched random disorder can lead to a softening of the transition \cite{Imry}.
Recently,  the predicted softening effect at first-order phase transitions
has been confirmed for 3D q-state Potts models with $q\geq3 $ using 
Monte Carlo \cite{balestro,chatelain1,chatelain2} and high temperature series
expansion \cite{M} techniques.
The overall picture is even better in two dimensions ($2D$) where several models
with $\alpha_{pure}>0$ \cite{D,Ma,Gj,SW} and the marginal ($\alpha_{pure}=0$) 
\cite{25,28,aarao,puli,puli2} have been investigated.

In this paper we study another type of quenched random disorder,
namely $connectivity$ $disorder$, a generic property of random lattices
whose local coordination number varies randomly from site to site.
Specifically, we consider $2D$ Poissonian random lattices of Voronoi-Delaunay type, 
and performed an extensive computer simulation study of a Blume-Capel model.
We concentrated on the close vicinity of the transition point
and applied finite-size scaling (FSS) techniques to extract the exponents and the
``renormalized charges" $U_{2}^{*}$ and $U_{4}^{*}$. To achieve the desired accuracy of the data
in reasonable computer time we applied the single-cluster hybrid algorithm \cite{Pla} to update 
the spins and furthermore made extensively use of the re-weighting technique \cite{his}.
Previous studies of connectivity disorder focusing mainly on $2D$ lattices have been realized 
by Monte Carlo simulations of $q$-state Potts models on quenched random lattices of 
Voronoi-Delaunay type for $q=2$ \cite{janke,janke1,FWSL}, $q=3$ \cite{FWSL1} and $q=8$ 
\cite{janke2,FWSL2}. In particular, it has been shown that for  $q=2$ 
\cite{janke,janke1,FWSL} and $q=3$  \cite{FWSL1} the critical
exponents are the same as those for the  model on a regular $2D$ lattice.
This is indeed a surprising result since the relevance criterion of the Delaunay triangulations 
reduces to the well known Harris criterion such that disorder of this type should be relevant
for any model with positive specific heat exponent \cite{JankeWeigel}. This means that  
for $q=3$, where  $\alpha_{pure}>0$,  one would expect a different universality class. 
On the other hand, for the present spin-3/2 model, where  $\alpha_{pure}=0$, 
we  show that the exponents indeed change in the  Voronoi-Delaunay lattice 
type, turning out the situation still more bizarre . In the next section we present the 
model and the simulation background. The results and
conclusions are discussed in the last section.

 \section{Model and  Simulation}

The Voronoi construction or tessellation for a given set of points
in the plane is defined as follows \cite{christ}. Initially, for each point one
determines the polygonal cell consisting of the region of space
nearer to that point than any other point. Then one considers that the
two cells are neighboring when they possess an extremity in
common. From the Voronoi tessellation the dual
lattice can be obtained  by the following procedure:
$(a)$ when two cells are neighbors, a link is placed between the
two points located in the cells;
$(b)$ From the links one obtains the triangulation of space
that is called the Delaunay lattice;
$(c)$ The Delaunay lattice is dual to the Voronoi tessellation in
the sense that points corresponding to cells link to edges, and
triangles to the vertices of the Voronoi tessellation.

We consider now the two-dimensional spin-$3/2$ Blume-Capel model on this
Poissonian random lattice. The  Blume-Capel Model is a generalization
of the standard Ising model \cite{kobe} and was originally proposed for spin-$1$ to account for 
first-order phase transition in magnetic systems \cite{BC,BC2}. The Hamiltonian
can be written as
\begin{equation}
H=-J\sum_{<i,j>}S_{i}S_{j}+\Delta\sum_{i}S_{i}^{2},
\end{equation}
where the first sum runs over all nearest-neighbor pairs of sites
(points in the Voronoi construction) and the spin-$3/2$ variables $S_{i}$ assume values 
$\pm 3/2,\pm 1/2$. In eq. (1) $J$ is the exchange coupling and $\Delta$ is the single ion
anisotropy parameter. The  second sum is 
taken over the $N$ spins on a $D$-dimensional lattice. 
The case where $S=1$ has been extensively studied by several approximate techniques
in two- and three-dimensions and its phase diagram is well established 
\cite{BC,BC2,BC3,BC4,BC5,BC6,BC7}. The case $S>1$ has also been investigated according
to several procedures \cite{BC8,BC9,BC10,BC11,BC12,BC13,landaupla}. 

The simulations have been performed for $\Delta=0$, which is the simplest case, on  
different lattice sizes  comprising a number $N=1000,2000,4000,8000$, $16000$ and $32000$ 
of sites. For simplicity,  the length of the system is defined here in terms
of the size of a regular lattice $L=N^{1/2}$.
For each system size quenched averages over the connectivity disorder are 
approximated by averaging over $R=100$ ($N=1000$ to $4000$), $R=50$ ($N=8000$) and $R=25$ 
($N=16000$ and $32000$) independent realizations. For each
simulation we have started with a uniform configuration of spins
(the results are however independent of the initial configuration). 
We ran $2.52\times10^{6}$ Monte Carlo steps (MCS) per spin with $1.2\times10^{5}$ configurations 
discarded for thermalization using the ``perfect" random-number generator \cite{nu}.
We have employed the hybrid algorithm \cite{Pla} where we included $n$ Wolff clusters 
(here $n=5$) intercalated by one Metropolis single-spin flip
sweep. This algorithm has been shown to be quite effective for spin-$3/2$ models \cite{Pla}.
For every $12$th MCS, the energy per spin,
$e=E/N$, and magnetization per spin, $m=\sum_{i}S_{i}/N$, were
measure and recorded in a time series file.

From the series of the energy measurements we can compute, by re-weighting over a controllable
temperature interval $\Delta T$, the average energy and specific heat
%
\begin{equation}
 u(K)=[<E>]_{av}/N,
\end{equation}
\begin{equation}
 C(K)=K^{2}N[<e^{2}>-<e>^{2}]_{av},
\end{equation}
%
where $K=J/k_BT$, with $J=1$, and $k_B$ is the Boltzmann constant. In the above equations
$<...>$ stands for  thermodynamic averages and $[...]_{av}$
for  averages over the different realizations.
Similarly, we can derive from the magnetization measurements
the average magnetization, the susceptibility, and the magnetic
cumulants,
\begin{equation}
 m(K)=[<|m|>]_{av},
\end{equation}
\begin{equation}
 \chi(K)=KN[<m^{2}>-<|m|>^{2}]_{av},
\end{equation}
\begin{equation}
 U_{2}(K)=[1-\frac{<m^{2}>}{3<|m|>^{2}}]_{av},
\end{equation}
\begin{equation}
 U_{4}(K)=[1-\frac{<m^{4}>}{3<|m|>^{2}}]_{av}.
\end{equation}
Further useful quantities involving both the energy and magnetization are
their derivatives
\begin{equation}
 \frac{d[<|m|>]_{av}}{dK}=[<|m|E>-<|m|><E>]_{av},
\end{equation}
\begin{equation}
 \frac{d\ln[<|m|>]_{av}}{dK}=[\frac{<|m|E>}{<|m|>}-<E>]_{av},
\end{equation}
\begin{equation}
 \frac{d\ln[<|m^{2}|>]_{av}}{dK}=[\frac{<|m^{2}|E>}{<|m^{2}|>}-<E>]_{av}.
\end{equation}
In the infinite-volume limit these quantities exhibit singularities at
the transition point. In finite systems the singularities are smeared
out and scale in the critical region according to
\begin{equation}
 C=C_{reg}+L^{-\alpha/\nu}f_{C}(x)[1+...],
\end{equation}
\begin{equation}
 [<|m|>]_{av}=L^{-\beta/\nu}f_{m}(x)[1+...],
\end{equation}
\begin{equation}
 \chi=L^{-\gamma/\nu}f_{\chi}(x)[1+...],
\end{equation}
\begin{equation}
 \frac{d \ln[<|m|^{p}>]_{av}}{dK}=L^{1/\nu}f_{p}(x)[1+...],
\end{equation}
where $C_{reg}$ is a regular background term, 
 $\nu$, $\alpha$, $\beta$, and $\gamma$ are the usual critical
exponents, and $f_{i}(x)$ are FSS functions with
$ x=(K-K_{c})L^{1/\nu}$
being the scaling variable, and the brackets $[1+...]$ indicate
corrections-to-scaling terms. We calculated the error bars from the fluctuations 
among the different realizations. Note that these errors contain both, the average
thermodynamic error for a given realization and the theoretical
variance for infinitely accurate thermodynamic averages which are
caused by the variation of the quenched, random geometry of the lattices.

\section{Results and conclusion}

By applying standard re-weighting techniques to each of the $R$ time-series data
we first determined the temperature dependence of $C_{i}(K)$, $\chi_{i}(K)$,..., $i=1$,...,$R$,
in the neighborhood of the simulation point $K_{0}$.
Once the temperature dependence is known for each realization, we can easily compute
the disorder average, e.g., $C(K)=\sum^{R}_{i=1}C_{i}(K)/R$, and then determine the maxima 
of the averaged quantities, e.g., $C_{max}(K_{max})=max_{K}C(K)$. The variable $R$ 
represents the number of replicas in our simulations.

\begin{figure}[ht]
\includegraphics[clip,angle=0,width=8.5cm]{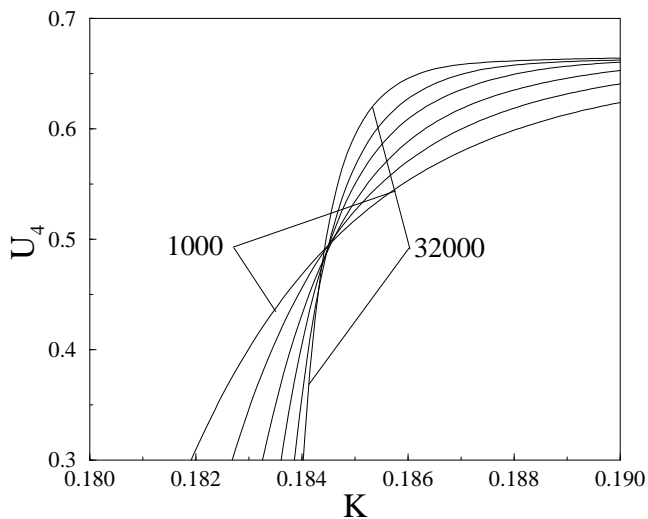}
\caption{\label{cum}
Fourth-order Binder cumulant as a function of $K$ for several values of the system 
size $N=1000, 2000, 4000, 8000, 16000$ and $ 32000$.
}
\end{figure}

In  order to estimate the critical temperature we calculate the second and fourth-order 
Binder cumulants given by  eqs. (7) and (8), respectively. It is well known that these
quantities are independent of the system size and should intercept at the critical 
temperature \cite{binder}. In Fig. \ref{cum} the fourth-order
Binder cumulant is shown as a function of the $K$ for several values of $N$. Taking the 
largest lattices we have  $K_{c}= 0.1844(1)$. To estimate $U^{*}_{4}$ we note that it
varies little at  $K_{c}$ 
so we have $U^{*}_{4}= 0.482(6)$. From the second-order
cumulant we similarly get $K_{c}= 0.1845(1)$ and $U^{*}_{2}= 0.579(8)$. One can see that 
the agreement of the critical temperature is quite good and  $U^{*}_{4}$ is definitely
far from the universal value $U^{*}_{4}\sim 0.61$  for the same model on the regular 
$2D$ lattice.
%
%
%
%

\begin{figure}[ht]
\includegraphics[clip,angle=0,width=8.5cm]{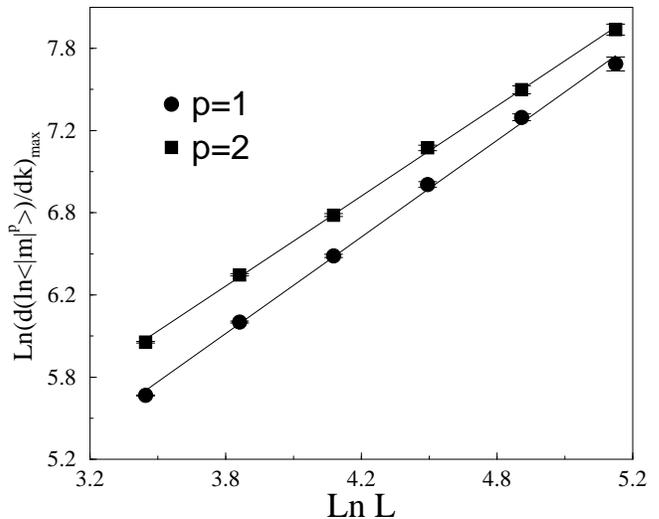}
\caption{\label{expnu}
Log-log plot of the maxima of the logarithmic derivative 
$\frac{d \ln[<|m|^{p}>]}{dK}$ versus  the lattice size $L=N^{1/2}$ for 
$p=1$ (circle) and $p=2$ (square). The solid lines are the best linear fits. 
}
\end{figure}

The correlation length exponent can be estimated from the derivatives given by
eq. (15). Figure \ref{expnu} shows the maxima of the logarithm derivatives as a 
function of the logarithm of the lattice size $L$ for $p=1$ and $p=2$. From the linear
fitting one gets $\nu=0.85(2)$ ($p=1$) and  $\nu=0.917(8)$ ( $p=2$), which is again
different from the regular lattice exponent $\nu=1$.
%
%
%

\begin{figure}[ht]
\includegraphics[clip,angle=-90,width=8.5cm]{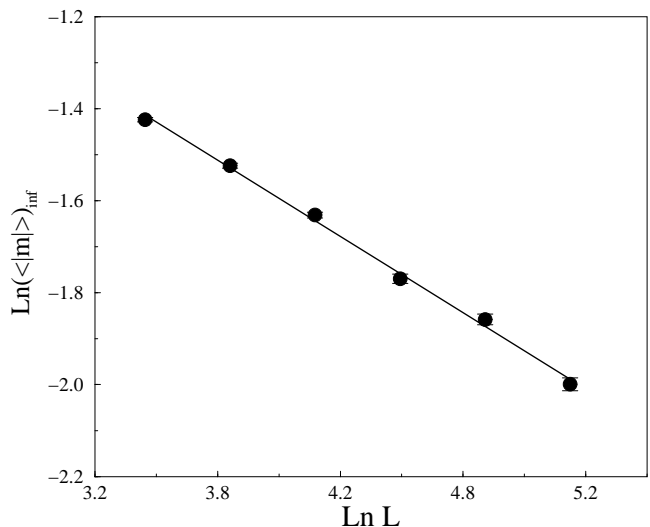}
\caption{\label{mag}
Plot of the logarithm of the modulus of the magnetization at the
inflection point as a function of the logarithm of  $L=N^{1/2}$. The solid line is 
the best linear fit.
}
\end{figure}

\begin{figure}[ht]
\includegraphics[clip,angle=0,width=7.9cm]{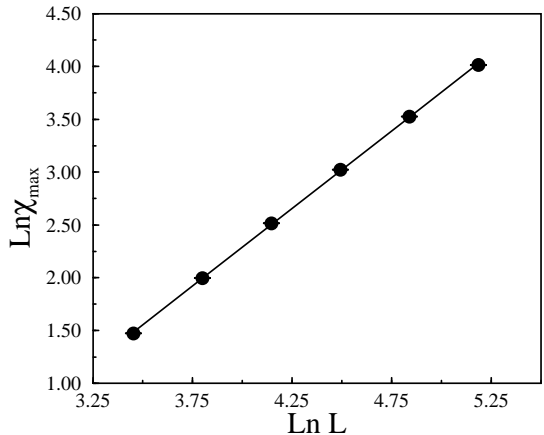}
\caption{\label{sus}
 Log-log plot of the susceptibility maxima $\chi_{max}$ as a function of
the logarithm of  $L=N^{1/2}$. The solid line is the best linear fit.
}
\end{figure}

In order to go further in our analysis we also computed the modulus of the magnetization
at the inflection point and the maximum of the magnetic susceptibility. The logarithm of
these quantities as a function of the logarithm of  $L$ are presented in Figures  
\ref{mag} and \ref{sus},
respectively. A linear fit of these data gives  $\beta/\nu=0.331(9)$ from the magnetization
and  $\gamma/\nu=1.467(9)$ from the susceptibility which should be compared to
 $\beta/\nu=0.125$ and  $\gamma/\nu=1.75$ obtained for a regular $2D$ lattice.

%
%
%
%
%
%
%

\begin{figure}[ht]
\includegraphics[clip,angle=0,width=8.0cm]{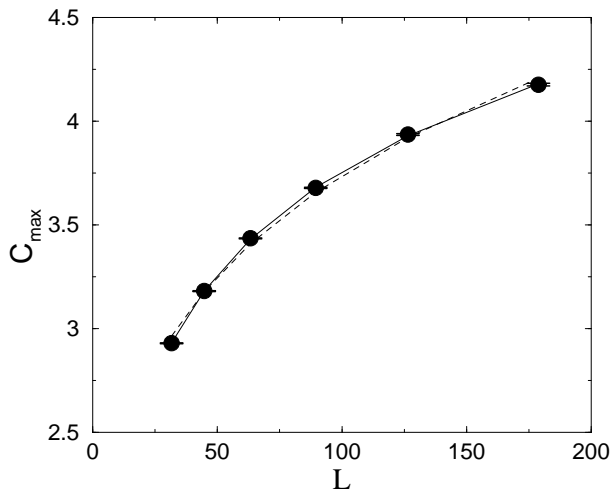}
\caption{\label{sph}
Specific heat maxima  $C_{max}$ as function of
$L=N^{1/2}$. The solid line is the best fit to an $\alpha\sim0$ (Log) Ansatz and
the dashed line to a power law Ansatz.
}
\end{figure}

The specific heat can also be analysed in this case but, as it happens in other
models \cite{janke1,FWSL1}, we cannot find a clear unambiguous support for a definite
scaling. Figure \ref{sph} shows the maximum of the specific heat  $C_{max}$  
as a function of $L$.  Least-squares fits to a logarithmic Ansatz 
$C_{max}=B_{0}+B_{1}\ln L$
give $B_{0}=0.44(6)$, $B_{1}=0.72(1)$ and is shown by the full line 
in figure \ref{sph}. The dashed line in this figure corresponds  to a pure 
power-law Ansatz, $C_{max}=cL^{\alpha/\nu}$ with $c=1.475(5)$
and $\alpha/\nu=0.202(5)$. From these results one can slightly see a better
agreement with the logarithmic Ansatz.
%
%
%

Thus, from the above results, there is a strong indication that the spin-3/2 Blume-Capel
model on a Voronai lattice is in a different universality class than its regular lattice 
counterpart. This poses, in addition to the $q=3$ Potts model in two dimensions, 
and taking into account the extensive study done Janke and Weigel on the Harris-Luck
criterion for random lattices \cite{JankeWeigel},
another  open question to be answered in more general terms.


\begin{thebibliography}{20}
\bibitem{puli} P. H. L. Martins and J. A. Plascak, to be published. In this work the use of
 the probability distribution function is used to show that the universality class of the $2D$
 diluted Ising model is in fact the same as that of the pure model.
\end{thebibliography}
\end{document}